\documentclass[mathleft
]{an}
\usepackage{graphicx}
\usepackage{times}
\usepackage{caption}
\usepackage{subcaption}
\begin{document}
\Pagespan{999}{}
\Yearpublication{?}%
\Yearsubmission{?}%
\Month{?}%
\Volume{?}%
\Issue{?}%

\title{Phase statistics of the WMAP 7 year data}

\author{A. Kov\'acs\inst{1,3}, I. Szapudi\inst{2,3} and Z. Frei\inst{1,3}\newline }
\titlerunning{Phase statistics of the WMAP 7 year data}
\authorrunning{A. Kov\'acs, I. Szapudi \& Z. Frei}
\institute{
E\"otv\"os University, Institute of Physics
\and 
University of Hawaii, Institute for Astronomy
\and
MTA-ELTE EIRSA "Lendulet" Astrophysics Research Group}

\accepted{2013}
\publonline{2013}


\abstract{We performed a comprehensive statistical analysis using complex phases of the $a_{lm}$ coefficients computed from the most recent data of the Wilkinson Microwave Anisotropy Probe (WMAP). Our aim was to confirm or constrain the presence of non-Gaussianities in the data. We found phase correlations - that suggest non-Gaussianity - at high-$l$ in $a_{lm}$ coefficients by applying various statistical tests. Most of all, we detected a non-Gaussian signal reaching a significance of 4.7$\sigma$ using random walk statistics and simulations. However, our conclusion is that the non-Gaussian behavior is due to contamination from galactic foregrounds that show up in small scales only. When masked out the contaminated regions, we found no significant non-Gaussianity. Furthermore, we constrained the $f_{\rm NL}$ parameter using CMB simulations that mimic primordial non-Gaussianity. Our estimate is $f_{\rm NL}=40\pm200$, in agreement with previous measurements and inflationary expectations.}

\maketitle

\section{Introduction}

Measurements of Cosmic Microwave Background (CMB) fluctuations and Large-Scale Structure are key science projects in the era of precision cosmology (Komatsu et al. 2003). Many cosmological theories are based on the idea that the galaxy distribution we observe today has grown from small initial density perturbations (Coles \& Chiang 2000) and the inflationary epoch played a crucial role. According to the simplest theory of inflation (Guth 1981), early quantum fluctuations are characterized by a statistically homogeneous, Gaussian random field (Bardeen et al. 1986). Perturbations are believed to be imprinted on the CMB at small angular variations (Gawiser \& Silk 2000). Consequently, they should have similar statistical properties. This similarity makes the CMB a useful tool to test the Gaussianity of the early Universe. However, many inflationary models predict some level of primordial non-Gaussianity generated during or shortly after inflation (Guth \& Pi 1985, Bartolo \& Mattarese \& Giotto 2010). These various mechanisms for the generation of cosmological density perturbations can be constrained with measurements of non-Gaussianities.

Several interesting variants of non-Gaussian properties have been claimed in the context of the CMB (Bennett et al. 2011). Low quadrupole (Efstathiou 2003), parity asymmetry (Bernui 2008; Kim \& Naselsky 2010), Cold Spot (Cruz et al. 2005; Vielva 2010), and the Axis of Evil (Land \& Magueijo 2005, 2007) are well-studied anomalies related to CMB non-Gaussianities, especially on large scales. A special family of statistical tests uses complex phases as an indicator of deviation from randomness (Coles \& Chiang 2000, Chiang \& Coles 2000; Watts \& Coles 2003). Phase mapping techniques (Chiang et al. 2003; Chiang \& Coles \& Naselsky 2002; Chiang \& Naselsky \& Coles 2004, Chiang \& Naselsky 2007), auto- and cross-correlations (Chiang \& Naselsky 2006), random walk tests (Stannard \& Coles 2005; Hansen et al. 2011), and even topological constraints (Dineen \& Coles 2003; Sung \& Short \&  Coles 2011) have been performed to uncover the properties of the CMB phases. Most of these studies diagnosed some form of non-Gaussian features in WMAP's earlier data releases. In this paper we examine the latest WMAP dataset with similar statistical tools, i.e. using complex phases.

The structure of this paper is as follows. In Section 2 we describe the data we used together with mathematical tools and statistical tests. We introduce our results of the non-Gaussianity analysis through examples in Section 3. Finally, we present our conclusions in Section 4 and discuss the meaning and interpretation of our findings.
\section{Data and methods}
We downloaded the CMB data from the LAMBDA website (http://lambda.gsfc.nasa.gov). We used WMAP's 7 year Internal Linear Combination (ILC) map, which was formed by a weighted linear combination of five different, smoothed temperature maps (Jarosik et al. 2011).

The weights were chosen to emphasize CMB anisotropies while minimizing the galactic foreground contribution.
The statistical characterization of temperature fluctuations can be expressed as a sum over spherical harmonics (Coles et al. 2004):
\begin{equation}
\frac{\Delta T}{T} (\theta,\phi) = \sum_{l=0}^{\infty} \sum_{m=-l}^{l} a_{lm} Y_{lm}(\theta,\phi)
\label{eq1}
\end{equation}
where complex $a_{lm}$ coefficients can be given by
\begin{equation}
a_{lm} = |a_{lm}| \cdot \exp{(i\phi_{lm})}
\label{eq2}
\end{equation}
We restrict our investigations to positive $m$ values, since $a_{lm}$'s have special symmetry relations (Coles et al. 2004).

Following Coles et al. (2004) we define a phase difference as
\begin{equation}
\Delta \phi (l,m) = \phi_{l+1,m} - \phi_{l,m}
\label{eq3}
\end{equation}
The usage of phase differences is preferable because of rotational properties of the $a_{lm}$ coefficients (Chiang et al. 2003). We use $\Delta \phi (l,m)$ phase differences to carry out our tests.

If the CMB is a realization of a homogeneous and isotropic Gaussian random field, real and imaginary parts of the $a_{lm}$ are independently distributed according to a Gaussian probability density. Therefore, $|a_{lm}|$ amplitudes are Rayleigh distributed, phases and phase differences are uniformly random on the interval $[-\pi,\pi]$ (Coles et al. 2004). Superposing many Fourier modes with random phases a Gaussian field is created. Phase statistics offer a convenient opportunity to test Gaussianity, statistical isotropy, and homogenity.

\subsection{Kuiper tests}

We use Kuiper's statistic  - that is a generalization of the Kolmogorov-Smirnov test  - to test the random phase hypothesis and analyze circular data (Kuiper 1960). The standard Kolmogorov-Smirnov statistic probes the maximum distance of the cumulative probability distribution against theoretical expectations. Kuiper's method introduces a statistic, V, obtained from the data: firstly, angles are sorted into ascending order, to give the set {$\theta_1,...,\theta_n$}.  Each angle $\theta_i$ is divided by $2\pi$ to give a set of variables $X_i$, where i = 1...n. From the set of $X_i$ we derive two sets $S_{n+}$ and $S_{n-}$ where
\begin{equation}
S_{n+} = max \left\{ \frac{1}{n} - X_1, \frac{2}{n} - X_2, ..., 1 - X_n \right\}
\end{equation}
and
\begin{equation}
S_{n-}=   max  \left\{  X_1, X_2-\frac{1}{n}, ..., X_n - \frac{n-1}{n}  \right\}
\label{eqsp}
\end{equation}
Kuiper's statistic, V, is then defined as
\begin{equation}
V=(S_{n+}+S_{n-}) \cdot \left(\sqrt{n}+0.155+\frac{0.24}{\sqrt{n}}\right)
\label{eqV}
\end{equation}
Large values of V indicate a distribution that is not fairly uniform (i.e. low Kuiper's $p$-value), while low values mean that the distribution is more regular.

\subsection{Random walk statistics}

The set of phase angles can be thought of as steps in a random walk (RW) on the complex plane, a structure that can be easily visualized and has well-known statistical properties (Stannard \& Coles 2005). We used a representation with unit step-size and the turning angle was characterized by the phases of the $a_{lm}$ coefficients. Provided that the angles are independent and uniformly random on the interval  $[-\pi,\pi]$
, the total displacement after $N$ steps is given by
\begin{equation}
R_N^2 = X_N^2 + Y_N^2
\label{eqsp}
\end{equation}
where $R_N$ is Rayleigh distributed (Pearson 1905; Rayleigh 1905). Moreover, $X_N$ and $Y_N$ are independently random.

On one hand, the actual traveled distance is computed in the case of the CMB data, on the other, it is compared to an expected value, $\mu$ and standard deviation, $\sigma$ that can be obtained from simulations. Using \texttt{synfast} we generated 1000 random CMB maps with $\Lambda CDM$ cosmological parameters (Jarosik et al. 2011). We obtained the distributions of the final positions after the walks using simulations, then calculated mean values and variances. A Rayleigh probability density is given by 
\begin{equation}
f(x,\sigma_0) = \frac{x}{\sigma_0^2} e^{-x^2/2\sigma_0^2}
\label{eqR}
\end{equation}
An estimator of the $\sigma_0$ parameter is expressed by
\begin{equation}
\hat \sigma_0 \approx \sqrt{\frac{1}{2N}\sum_{i=1}^{N}x_i^2}
\label{eq8}
\end{equation}
Evaluating parameters in Eq. \ref{eqM} a significance test can be performed to test Gaussianity.
\begin{equation}
\centering
\mu = \sigma_0 \sqrt{\frac{\pi}{2}}, ~~~ \sigma = \sqrt{\frac{4-\pi}{2}\sigma_0^2}
\label{eqM}
\end{equation}
\section{Results}

\begin{figure}[!h]
\centering
\includegraphics[width=75mm,height=55mm]{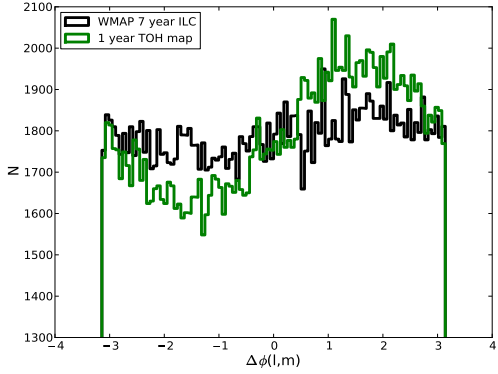}
\caption{Distribution of phase differences is presented with multipoles up to $l_{\rm max}=600$. The level of non-uniformity is different, WMAP's 7 year data shows weaker phase correlations.}
\label{hist}
\end{figure}

We computed $a_{lm}$'s using the \texttt{anafast} routine from HEALPIX (G\'orski et al. 2005). Firstly, we used a 1 year foreground cleaned CMB map (Tegmark \& de Oliveira-Costa \& Hamilton 2003) to reproduce former results in which non-Gaussianities were indicated (Chiang et al. 2003). We introduce our simplest results on Fig. \ref{hist} in terms of histograms. Visually, there are stronger phase-correlations in the 1 year data, although there are hints of deviations from uniformity using the 7 year WMAP product.

\begin{figure}[!h]
\centering
\includegraphics[width=80mm,height=60mm]{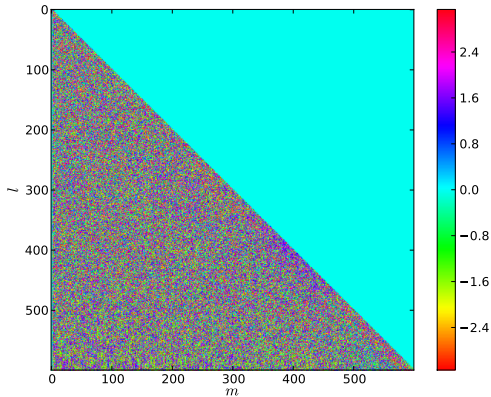}
\caption{Color-coded map of $\Delta \phi (l,m)$ phase differences is shown. As clearly visible, 
phases are correlated at $l \sim 400$, $l \approx m$. Also one finds non-random features at $l > 550$. Similar results were reported by Chiang et al. (2003).}
\label{tri}
\end{figure}

Histograms are not efficient to fully explore the statistical properties of the phases, since the source of the non-Gaussianity cannot be localized in the ($l,m$) space. We follow Chiang et al. (2003) to advance our techniques and use a color-coded representation of the phases. This result is shown on Fig. \ref{tri}. We identified phase correlation on small scales similarly to Chiang et al.'s result, although the majority of the map is random in a good approximation.
We obtained probabilities of measuring a V value (\ref{eqV}) as high as calculated from WMAP 1 year and 7 year datasets. Kuiper's $p$-value was $p_{K} <10^{-10}$ in both cases, i.e. these are highly significant detections of non-uniform features. However, the positive signal seems to come from the small scales only, which are rather dominated by noise (Larson et al. 2011). 

\begin{figure}[!h]
 \centering
\includegraphics[width=80mm,height=60mm]{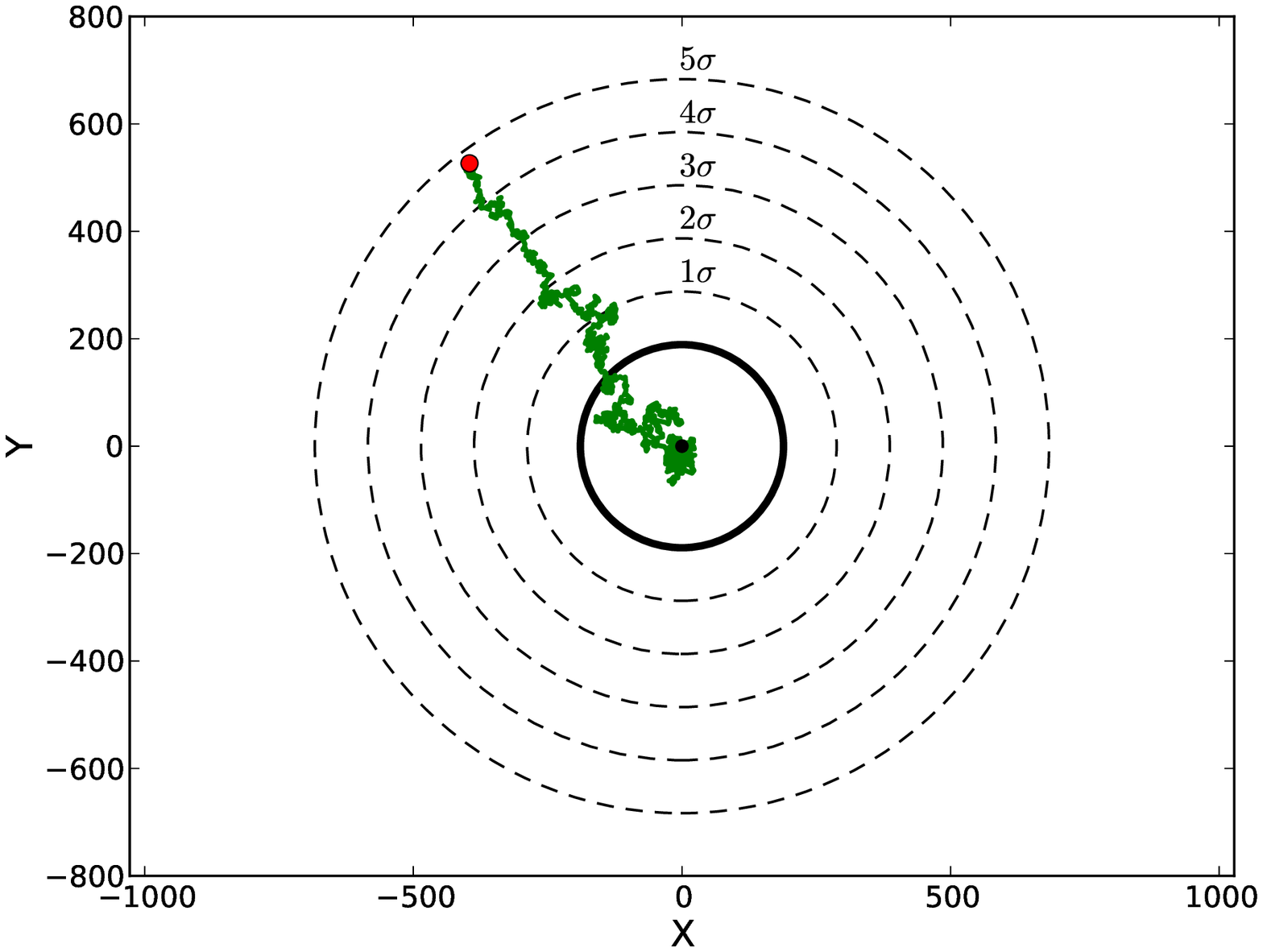}
\includegraphics[width=80mm,height=60mm]{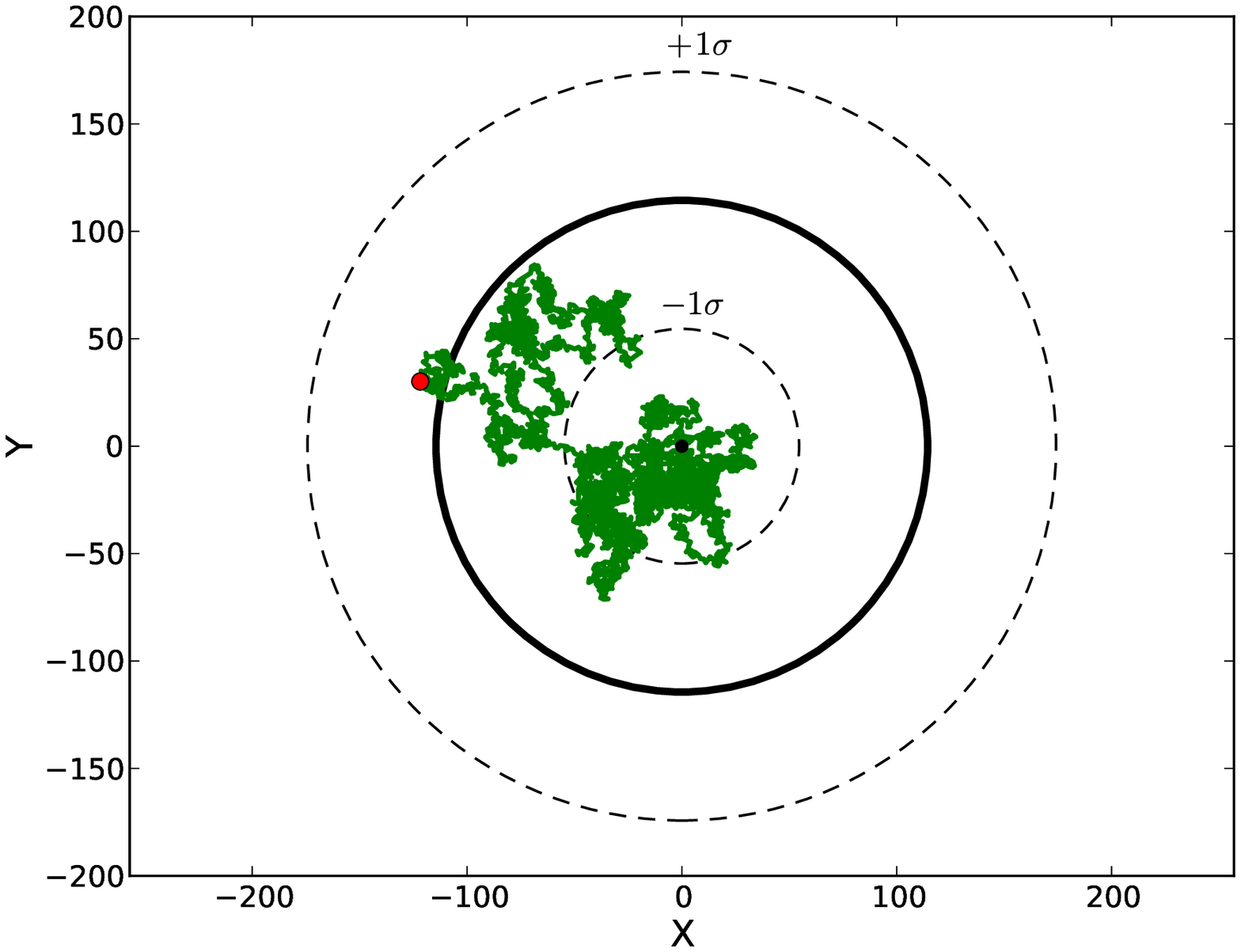}
\caption{Solid black circles show expected values of final distances after the walk, dashed circles show regions with varying significance. Top: $l_{\rm max}=300$, where $\mu=189.1, \sigma=98.8$. Bottom: $l_{\rm max}$=200, where $\mu=114.4$, $\sigma=59.8$}.
\label{walk1}
\end{figure}

Next, we analyzed phases at the regions $l_{\rm max}<300$ applying RWs. The walk was built by placing $\Delta \phi (l,m)$ into ascending order primarily by $l$, secondarily by $m$. We measured a 4.7$\sigma$ difference from the expected value. It is shown on Fig. \ref{walk1} that after the 30-40\% of the steps the walk turned to be non-random and ended very far from the expected regions.
If we restrict our tests below $l_{\rm max}=200$ the results are different. The highly significant deviation has disappeared as it is displayed on the bottom of Fig. \ref{walk1}.
\section{Discussion \& Conclusions}

We adopted a set of statistical tests to carry out tests of non-Gaussianity using WMAP's latest data. We learned from histograms and color-coded plots that the knowledge of non-Gaussianities is weak without knowing its origin in ($l,m$) space. Results of Kuiper's test indicated some form of non-Gaussianity, but only at $l$-ranges where noise strongly affects the CMB measurement.

However, we detected non-Gaussian features at $200<l<300$ with RWs. Noise is not meaningful on these scales, but possible remnant galactic foregrounds can cause an uncertainty factor. We used WMAP's original Q, V and W temperature maps together with their foreground reduced versions to analyze this effect.
\begin{table}[!h]
\centering
\caption{We tested both raw and foreground removed (FR) versions of the CMB temperature maps. Best possible constraints of the $f_{\rm NL}$ parameter are shown in the last column. We used positive values of $f_{\rm NL}$ assuming its effects are symmetric.\newline}
\label{tlab}
\begin{tabular}{@{}lcccccccccc}
Map type & Mask & $R_N$ & Significance & $f_{\rm NL}$\\ 
\hline
WMAP ILC & - & 658.5 & 4.7$\sigma$ & $>1000$ \\
WMAP ILC & + & 266.5 & 0.8$\sigma$ & $100\pm200$\\
\hline
WMAP Q & + & 254.5 & 0.6$\sigma$ & $60\pm200$  \\
WMAP V & + & 218.9 & 0.3$\sigma$ & $20\pm200$ \\
WMAP W & + & 190.9 & 0.0$\sigma$ & $0\pm200$ \\
\hline
WMAP FR Q & + & 130.2 & -0.6$\sigma$ & $50\pm200$ \\
WMAP FR V & + & 166.2 & -0.2$\sigma$ & $20\pm200$ \\
WMAP FR W & + & 141.8 & -0.5$\sigma$ & $40\pm200$ \\
\hline
\label{tabla}
\end{tabular}
\end{table}
These sky maps are contaminated by point sources and our galaxy, but using the WMAP team's mask we were able to cut out these regions. The orthogonality of spherical harmonics is no longer preserved with this procedure, therefore we consider this as an exploratory test and interpret the results with care. We cannot report any positive non-Gaussian signals performing the previous RW statistics with these masked CMB maps. Our findings are summarized in Table \ref{tabla}.

\begin{figure}[!h]
\centering
\includegraphics[width=80mm,height=55mm]{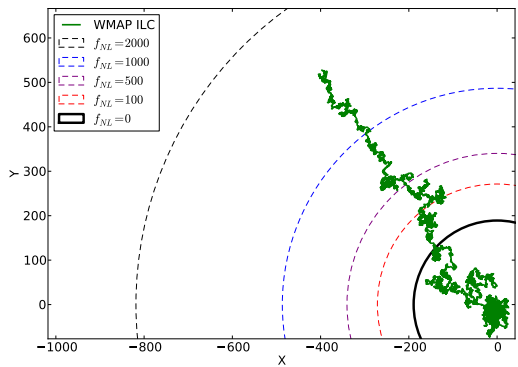}
\caption{The WMAP ILC, $l_{\rm max}=300$ random walk is shown with expected values of $R_N$ distances using different non-Gaussian CMB simulations.}
\label{fnl}
\end{figure}

Finally, we modeled the possible primordial origin of the signal we found. We used non-Gaussian CMB simulations and computed $a_{lm} = a_{lm}^{L} + f_{\rm NL}\cdot a_{lm}^{\rm NL}$, where $f_{\rm NL}$ can be any desired level of non-Gaussianity (Elsner \& Wandelt 2005). We measured expected values of $R_N$ distances varying the $f_{\rm NL}$ value. $f_{\rm NL}>1000$ could explain the properties of the RW, according to Fig. \ref{fnl}. However, previous measurements constrained $f_{\rm NL}=32\pm21$ (Komatsu et al. 2011). This outcome suggests that the detected non-Gaussian properties cannot be explained by $f_{\rm NL}$-like primordial effects. We calculated the possible $f_{\rm NL}$ constraints with $1\sigma$ error bars using our RW measurements, the average of our estimations is $f_{\rm NL}\approx40\pm200$. These outcomes are summarized in Table \ref{tabla}. We note that our tests are not optimized for $f_{\rm NL}$ non-Gaussianity constraints.

In the light of the performed tests, we interpret our results as a detection of residual foregrounds rather than cosmological defects. This is fully consistent with Gaussianity and previous findings (Chiang et al. 2003; Coles et al. 2004; Naselsky et al. 2005). A remnant contamination is still present in the ILC map, even after several improvements on the foreground cleaning. These artifacts are imprinted on the phase statistics we used in this work.

\acknowledgements
This research was supported by NASA grants NNX12AF83G and NNX10AD53G and the Polanyi program of the Hungarian National Office for the Research and Technology (NKTH), AK and ZF acknowledges support from OTKA through grant no. 101666.

\end{document}